\documentclass{article}
\usepackage{amssymb, amsmath, amsfonts, amsthm}
\usepackage{authblk} 
\usepackage{enumerate}
\usepackage{graphicx}
\usepackage{pdfpages}
\usepackage{multicol}
\usepackage{float}
\usepackage{caption}
\usepackage{subcaption}
\usepackage{setspace}
\usepackage[title]{appendix}
\floatstyle{plaintop}
\restylefloat{table}
\usepackage[tableposition=top]{caption}
\usepackage[greek, english]{babel}

\newcommand{\rel}{\operatorname{rel}}
\newcommand{\tsup}{\operatorname{tsup}}
\newcommand{\Ji}{\operatorname{Ji}}

\usepackage[backend=biber, style=numeric]
{biblatex}
\renewbibmacro{in:}{}
\DeclareFieldFormat{pages}{#1}
\DeclareFieldFormat[article, incollection, book, inproceedings,     unpublished]{title}{#1}
\theoremstyle{plain}% default
\newtheorem{thm}{Theorem}

\theoremstyle{definition}

\newtheorem{exam}{Example}

\title{Space efficient implementation of hypergraph dualization in the $D$-basis algorithm}
\author[1]{Skylar Homan}
\affil[1]{Programs Department, AMS}
\author[2]{Mikhail Khlystun}
\affil[2]{Star HPC System Administrator, Hofstra University}
\author[3]{Anoop Krishnadas}
\affil[3]{School of Engineering, Hofstra University}
\author[4]{Kira Adaricheva\footnote{Corresponding author; email: Kira.Adaricheva@hofstra.edu}}
\affil[4]{Department of Mathematics, Hofstra University}

\date{}

\begin{document}
\maketitle

\begin{abstract} We present a new implementation of the $D$-basis algorithm called the \emph{Small Space} which considerably reduces the algorithm's memory usage for data analysis applications. The previous implementation delivers the complete set of implications that hold on the set of attributes of an input binary table. In the new version, the only output is the frequencies of attributes that appear in the antecedents of implications from the $D$-basis, with a fixed consequent attribute. Such frequencies, rather than the implications themselves, became the primary focus in analysis of datasets where the $D$-basis has been applied over the last decade. The $D$-basis employs a hypergraph dualization algorithm, and a dualization implementation known as \emph{Reverse Search} allows for the gradual computation of frequencies without the need for storing all discovered implications. We demonstrate the effectiveness of the Small Space implementation by comparing the runtimes and maximum memory usage of this new version with the current implementation. 
\end{abstract}


\section{Introduction}

In data mining, the retrieval and sorting of association rules is a research problem of considerable interest. Association rules uncover the relationships between the attributes of a set of objects recorded in a binary table.

This, for example, can be a medical data table, where the objects/rows are patients and the attributes/columns are indicators of some property, such as a genetic marker or record of a particular treatment, that may or may not hold in a patient: position $(r,c)$ in the table, in row $r$ and column $c$, is $1$ if property $c$ occurs in patient $r$; otherwise, it is $0$.

The algorithm known as \emph{Apriori} remains a common tool for extraction of association rules from binary data and was included in libraries of R and Microsoft office. One of the more recent surveys on association rules is in \cite{Bal2010}. 

One particular subset of association rules known as \emph{implications}, or rules of full confidence, merit particular attention in data mining, as well as being at the center of ongoing theoretical study (\cite{BDVG2016,Wild2017}).
From a practical point of view, implications are the strongest rules available on a given table because they hold true for any row of the table.
 
The retrieval of a set of implications as an equivalent representation of a binary table has been developed using ideas from Formal Concept Analysis (FCA); see \cite{Wil99}.
It is based on the fact that the table generates a unique closure operator on the set of its attributes, and thus can be associated with the lattice of the closed set of this operator, also known as the \emph{Galois lattice} or \emph{the concept lattice}. The standard approach in FCA is to build the concept lattice, then retrieve its set of implications, known as the \emph{canonical basis}.
The size of the concept lattice is exponential on the size of the table, and the retrieval of the canonical basis is co-NP complete \cite{BK2010}; thus, these algorithms are practical only for tables of small size.

A new approach was taken in \cite{AN17}, where the canonical basis of the concept lattice is replaced by the $D$-basis, a set of implications with a background in the study of free lattices \cite{FJN95}. The key difference compared to the canonical basis approach is that the computation of the $D$-basis does not require the whole concept lattice. Instead, the $D$-relation on the reduced set of attributes is used to find a portion of the $D$-basis associated with a particular consequent $t$ from the set of attributes. The crucial advantage is in the use of a \emph{hypergraph dualization} algorithm on the specific hypergraph associated with the attribute $t$ and the $D$-relation. The edges of the dual hypergraph are used to produce the output of the $D$-basis implications. This algorithm was dubbed the $D$-\emph{basis algorithm}.

We note that the hypergraph dualization was suggested as a tool for obtaining the \emph{canonical direct unit basis} in the number of earlier work \cite{Pfa02, DRB11,Sza14}. This basis unifies the multiple equivalent types of implicational bases discussed in \cite{Ber2010}. As shown in \cite{ANR13}, the $D$-basis is a subset of the canonical direct unit basis, and the cardinality gap between these two bases depend on the number of binary implications, i.e., implications $a\to b$, where $a,b$ are attributes. For example, one standard testing example of the table with 10 rows and 22 attributes, would have 13 binary implications, which brought to a considerable gap between two bases: about 800 implications in the $D$-basis and 1500 in the canonical direct basis. 

Since the $D$-basis approach employs hypergraph dualization, which is an algorithm with sub-exponential time complexity based on the size of the input and output,
the $D$-basis is very effective in terms of the time spent to generate any given implication in the output. On the other hand, it contains considerably more implications than the canonical basis, and thus may require a larger amount of space for its output.
The output shows this exponential tendency, per testing results done on a series of tables of increasing size with the density of ones in the middle range of 0.3-0.7
\cite{SCAN18}.

In the current project, a new approach was taken with respect to the usage of hypergraph dualization in the implementation of the $D$-basis algorithm.

Observe that in many applications, only the aggregate values related to
frequencies of attributes that appear in implications were needed, as opposed to requiring the complete list of implications; see \cite{stem2025, ABLS23, NCSLA21,}. These aggregate values can be updated every time a transversal is found during the hypergraph dualization stage of the algorithm. This task can be achieved using the \emph{Reverse Search} implementation of dualization described in \cite{MU14}.

The new implementation of the $D$-basis algorithm, called the \emph{Small Space}, makes use of the enumerative properies of Reverse Search to relocate the aggregate computation into the hypergraph dualization stage of the algorithm without affecting its output accuracy. This removes the need to allocate space for recording and transferring the list of implications between the stages of the algorithm.

In section 2, we introduce important concepts to provide for the background of the $D$-basis algorithm as it is formulated in Theorem \ref{D&dual} \cite{AN17}.
Section 3 gives an overview of the attribute ranking process of tabled data with respect to a set of implications, which was developed as part of the $D$-basis output in \cite{ANO15}. In particular, we demonstrate the key difference between the existing version and the new \emph{Small Space} version of the $D$-basis algorithm.
In Section 4, we provide test results of the Small Space algorithm side by side with the existing $D$-basis algorithm that demonstrate the space savings when the new version of the algorithm is used.

\section{Tables, closure systems, implicational bases and hypergraphs}

In this section we will follow a simple example to present the major concepts and the mathematical background of the $D$-basis algorithm. Many additional details can be found in \cite{AN17}. 

We will call $T=(U, A, R)$ a binary table if $A,U$ are finite sets and $R\subseteq U\times A$ a \emph{relation} between $U$ and $A$. $U$ is called a set of objects and $A$ a set of attributes.

\begin{exam}
Let $T=(U, A, R)$ be a binary table given in  Table \ref{Tab13*}. Here $U=\{1,2,\dots, 9\}$, $A=\{t, a_1,a_2, b_1, b_2, c\}$ and pairs from $R$ are marked by 1 in the table.

\begin{table}[!t]
\begin{center} 
\begin{tabular}{ |  c| c | c|c|c|c|c  |} 
  \hline
  $U\setminus A$ &  $t$ & $a_1$ &$a_2$ & $b_1$ & $b_2$ & $s$     \\
  \hline
 1  & 0 &  1  & 0 &  1 & 0 & 0 \\
  \hline
  2  & 1 &  0  & 0 &  1 & 1 & 0 
 \\
 \hline
  3  & 0 &  0  & 1 &  0 & 1 & 1 \\
 \hline
 4  & 0 &  0  & 0 &  1 & 1 & 0\\
 \hline
 5  & 1 &  1  & 1 &  1 & 1 & 0 \\
 \hline
 6  & 0 &  0  & 0 &  1 & 1 & 0\\
\hline
7  & 1 &  0  & 0 &  1 & 1 & 0\\
\hline
8  & 1 &  1  & 1 &  1 & 1 & 1\\ 
\hline
9  & 1 &  1  & 1 &  1 & 1 & 0 \\
\hline
\end{tabular}
\caption{Binary table with 6 attributes}
\label{Tab13*}
\end{center}
\end{table}

\end{exam}

For any subset $Y\subseteq A$ there exists $Z\subseteq U$ such that $z\in Z$ iff $(z,y)\in R$ for all $y\in Y$. $Z$ is called a \emph{support} of $Y$ and denoted $Z=sup(Y)$. This formally defines mapping $sup_A: 2^A\rightarrow 2^U$, a support function on $A$. Symmetrically, $sup_U$ can be defined on $U$. These functions are typically used in pattern mining in FCA, see \cite{GK01}. Their notation in FCA literature is given by a prime operator  $'$ , for both $sup_A$ and $sup_U$. For example, $Z=sup_A(Y)$ would be denoted $Z=Y'$.

\begin{exam}
In Table \ref{Tab13*}, $sup_A(\{a_2,b_2\})=\{3,5,8,9\}$ and $sup_U(\{6,7\}=\{b_1, b_2\}$.
\end{exam}

It is a standard and easy argument to show that $\phi_A: 2^A\rightarrow 2^A$, defined as $\phi_A(Y)=sup_U(sup_A(Y))$ for any $Y\in 2^A$, is a \emph{closure operator}, i.e., it is increasing, monotone and idempotent. Sets $\phi_A(Y)$, $Y\in 2^A$, form a \emph{closure system}, i.e., a set system with a family of subsets of $A$ closed under intersection and containing $A$ itself. Such a family always forms a lattice with respect to the inclusion order. This lattice $L(T)$ is called the \emph{Galois lattice}, or the \emph{concept lattice} of table $T=(U, A, R)$. We mention that the table is called \emph{reduced} when operator $\phi_A$ is \emph{standard}, i.e. satisfies the condition that $\phi_A(\{y\})\setminus \{y\}$ is closed for every $y\in A$. In such a case, set $A$ is in one-to-one correspondence with the set of join irreducible elements of lattice $L(T)$, which are exactly $\phi_A(y)$, $y\in A$. 

Note that one can symmetrically define the closure operator $\phi^*: 2^U\rightarrow 2^U$ on the set $U$ by setting $\phi^*(Z)=sup_A(sup_U(Z))$ for any $Z\in 2^U$. The resulting lattice of closed sets is $L^*(T)$, which is dual to $L(T)$. In particular, the set of join irreducible elements of $L^*(T)$ is in one-to one correspondence with $z\in U$. They will correspond to the meet irreducible element of $L(T)$. Note that $L^*(T)$ is a typical representation of the Galois lattice in FCA, i.e., the elements of $A$ are in one-to-one correspondence with the meet irreducible elements of the lattice.

One can follow monographs \cite{Birk40,Barb70, Wil99} on the topic. 

\begin{figure}[htbp] 
\begin{center}
\includegraphics[scale=0.4]{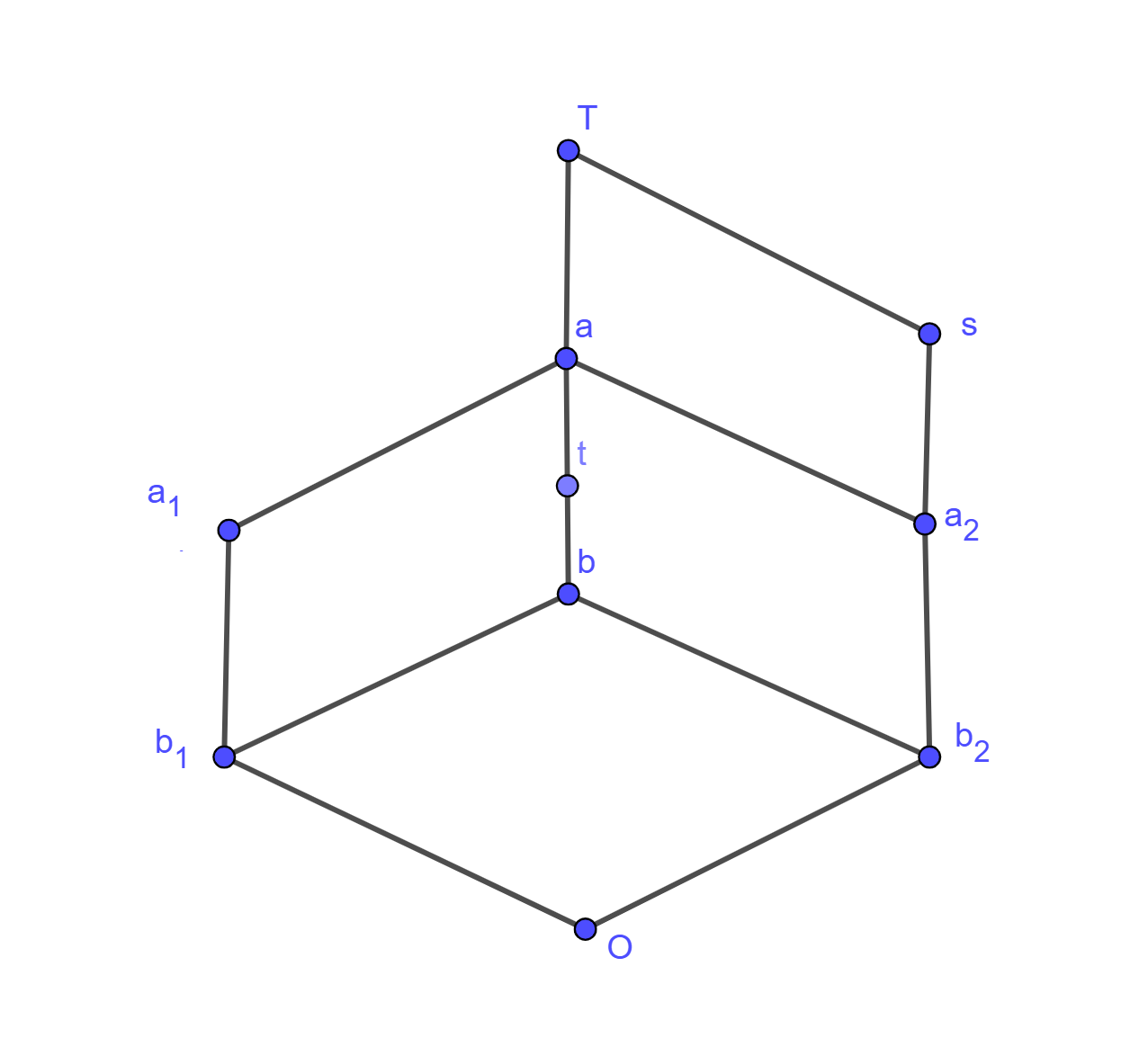}
\caption{The Galois lattice $L(T)$ of Table  \ref{Tab13*} }
\label{fig:lattice6}
\end{center}
\end{figure}

\begin{exam}
One can check that $\phi_A(\{t\})=\{t,b_1, b_2\}$ and $\phi^*(\{8,9\}=\{5,8,9\}$. The Hasse diagram of $L(T)$ is on Figure \ref{fig:lattice6}, where $x$ marks set $\phi_A(\{x\})$ for all $x\in A=\{a_1,a_2,b_1,b_2,s,t\}$.
\end{exam}

It is typical to retain partial information about a closure operator $\phi$ on set $A$ known as an \emph{implicational basis} (IB) so that the operator can be fully recovered from it. Every IB can be represented as a set of pairs $(X,Y)\subseteq 2^A\times 2^A$ such that $Y\subseteq \phi(X)$; more often, such pairs are written as an implication $X\rightarrow Y$.

\begin{exam}
It was established in the previous Example that $\phi_A(\{t\})=\{t,b_1, b_2\}$, which can be written as the implication $t\to b_1b_2$. Its meaning in Table $T$ is that in every row of the table where the entry for column $t$ is marked by 1, the entries for columns $b_1$ and $b_2$ in that row are also marked by 1.
\end{exam}

For any IB there is a unique corresponding closure operator, but there are multiple IBs for the same closure operator.  

A well known example of an IB is the \emph{canonical} basis, or Guigues-Duquenne basis (\cite{GD86}), $\Sigma_c=\{P\rightarrow \phi(P): P\subseteq A \text{ is a \emph{pseudo-closed} set}\}$. The canonical basis is known for having the minimal number of implications defining the closure operator. See \cite{Wild96} for details.

Before proceeding to the description of another IB known as the $D$-basis, it is necessary to introduce the concepts of a hypergraph and its dual.

A \emph{hypergraph} is a pair of a set and a family of subsets: $\mathcal{H}=\langle A, \mathcal{F}\rangle$, $\mathcal{F}\subseteq 2^A$.
A \emph{transversal} of the hypergraph is $Y\subseteq A$ such that $Y\cap F\not = \emptyset$ for every $F\in \mathcal{F}$. $Y$ is called \emph{minimal} if $Y\setminus y$ is not a transversal for every $y\in Y$. The hypergraph $\mathcal{H}^d=\langle A, \mathcal{T}\rangle$, where $\mathcal{T}$ is a set of minimal transversals of $\mathcal{H}=\langle A, \mathcal{F}\rangle$, is called the \emph{dual} of $\mathcal{H}$.

The hypergraph duality problem is to recognize, given two hypergraphs $\mathcal{G}$ and $\mathcal{H}$ on the same set of vertices, whether $\mathcal{G}=\mathcal{H}^d$. The process of generating $\mathcal{H}^d$ for a given hypergraph $\mathcal{H}$ is called \emph{dualization}.
Dualization is an enumeration algorithm, i.e. one that produces all required objects and does so without repetitions. In the case of dualizing a hypergraph $\mathcal{H}$, all minimal transversals of $\mathcal{H}$ have to be produced, and each must appear in the output only once. Many optimization problems reduce to the task of dualization of some hypergraph, which makes this problem very important in applied mathematics and computer science.

It turns out that the dualization of the hypergraph can be incorporated into the procedure of finding the $D$-basis.

The $D$-basis was introduced in \cite{ANR13} and belongs to the family of \emph{unit} bases, i.e. 
IBs comprised of implications $Y\rightarrow x$, with $Y\subseteq A$ and $x \in A$. While there is no technical difference between unit bases and arbitrary IBs, the focus of the unit implication is on the consequent element $x$ rather than the set $Y$ in the left side of implication. In particular, the $D$-basis was defined as the implicational equivalent of the \emph{join cover relations} known in lattice theory as the way to define lattices.

More precisely, every element of a finite lattice $L$ is a join of join irreducible elements, the set of which is denoted $\Ji(L)$, so the relation can be written in the form $x \leq \bigvee Y$, for $x\in \Ji(L)$ and $Y\subseteq \Ji(L)$. If $Y=\{y\}$, then this is the order relation $x\leq y$ for some elements $x,y\in \Ji(L)$, and these relations are included in any IB. Otherwise,  $Y$ is called a \emph{non-trivial minimal join cover} for $x$ if $x\not\leq y$ for $y\in Y$ and none of $y\in Y$ can be replaced by smaller elements in $L$ and still produce a cover for $x$. 

\begin{exam}
$t\leq a_1\vee s$ is a non-trivial join cover for $t$ in lattice $L(T)$ on Figure \ref{fig:lattice6}, but it is not a minimal cover, because $t\leq a_1\vee b_2$, for $b_2 < s$, and neither $a_1$ nor $b_2$ can be replaced by smaller elements to make a join cover for $t$.
\end{exam}

Every lattice $L$ with $A=\Ji(L)$ can be represented as a closure system on set $A$ with the closure operator $\phi$ defined by the IB comprising \emph{binary implications} $y\to x$ for all relations $x\leq y$, and  $Y\rightarrow x$ for minimal covers $x\leq \bigvee Y$ in $L$. This IB is called the $D$-basis \cite{ANR13}.

\begin{exam}
The $D$-basis for $L(T)$ on Figure \ref{fig:lattice6} consists of binary implications $a_1\to b_1$, $t\to b_1$, $t\to b_2$, $a_2\to b_2$, $s\to a_2$, and $s\to b_2$;
and non-binary implications $a_1b_2\to t$, $a_1b_2\to a_2$, $b_1a_2\to t$ and $b_1a_2\to a_1$.
\end{exam}

The $D$-relation on $\Ji(L)$ can be defined independently of the $D$-basis, but the following is a necessary and sufficient condition (\cite{FJN95}):\\

$xDy$ iff  $y\in Y$ for some non-binary implication $Y\to x$ in the $D$-basis. \\

Now we connect this with the $D$-relation as it is identified in a table $T$ with Galois lattice $L(T)$.

There exist additional relations $\uparrow, \downarrow, \updownarrow \subseteq U\times A$ 
that can be easily computed given $T=(U,A,R)$. Namely, $(w,x)\in \uparrow$ iff $x\not \in sup_U(w)$, but $x\in sup_U(z)$ for every $z\in U$ with $sup_U(w)\subset sup_U(z)$.

Symmetrically, $(w,x)\in \downarrow$ iff $w\not \in \sup_A(x)$, but $w\in \sup_A(y)$ for every $y\in A$ with $sup_A(x)\subset sup_A(y)$. Finally $\updownarrow = \uparrow \cap \downarrow$.

Many properties of the Galois lattice can be expressed in the language of these relations \cite{Wil99}. It is the case that $\uparrow, \downarrow, \updownarrow \subseteq (U\times A) \setminus R$, and the $D$-relation on set $A$ can be computed using $\uparrow, \downarrow$ (\cite{FJN95}):\\

For any $x,y\in A$, $xDy$ iff there exists $w\in U$ such that $(w,x)\in \uparrow$ and $(w,y)\in \downarrow$. \\

Further details can be found in \cite{AN17}. We denote $xD=\{y\in A: xDy\}$.

\begin{exam}\label{table}
In Table \ref{Tab13*}, $(3,t)\in \uparrow$ and $(3,b_1)\in \downarrow$, therefore, $tDb_1$. In fact, $tD=\{a_1, a_2, b_1, b_2\}$.
\end{exam}

Now we want to explore the possibility of finding the $D$-covers for one of elements in $A$. For example, given binary table $T=(U,A,R)$, suppose that one needs to find all possible implications $Y\to x$ for a fixed element $x\in A$ and some $Y\subseteq A$. In such a setting, element $x\in A$ is called a \emph{target}. It is easy to find trivial covers for $x$ directly in the table, so we focus on non-trivial ones.

This is a task suitable for a unit basis such as the $D$-basis, because it is equivalent to the task of finding all non-trivial minimal join covers $x \leq \bigvee Y$ in lattice $L(T)$ for that element $x$. (Recall that $x\in A$ corresponds to $\phi_A(x)\in L(T)$, so we can continue calling $\phi_A(x)$ as $x$.)

Note that since $x \leq \bigvee Y$, then $\bigvee Y$ is one of the elements in interval $[x,1_L]\subseteq L(T)$, while we need to pick elements for $Y$ in $L(T)\setminus [x,1_L]$. 

Denote $M(x)=\{M_1,\dots, M_k\}$ as the set of maximal elements in the poset $L(T)\setminus [x,1_L]$. Then $\bigvee Y \not \leq M_i$, so $Y\not \subseteq M_i$ for all $i\leq k$. Therefore, $Y$ should have a non-empty intersection with $A\setminus M_i$ for each $i\leq k$. Additionally, we want $Y$ to be a non-trivial minimal cover, i.e. $Y\subseteq xD$. Therefore, $Y$ should have a non-empty intersection with all sets $xD\setminus M_1, \dots, xD\setminus M_k$. Moreover, due to the definition of a minimal cover, $|Y|$ is minimal with this property. 

This makes $Y$ a \emph{minimal transversal} in hypergraph $\mathcal{H}(x)=\langle xD, \{xD\setminus M_1, \dots, xD\setminus M_k\}\rangle$.

The argument above is an outline of the following result in \cite{AN17}.

\begin{thm}\label{D&dual}
Let $T=(U,A,R)$ be a (reduced) table, $x\in A$ and $M(x)=\{M_1,\dots M_k\}$ be the set of maximal elements in $L(T)\setminus [x,1_L]$. Then for any non-binary implication $Y\to x$ in the $D$-basis of operator $\phi_A$, set $Y$ is a minimal transversal of the hypergraph $\mathcal{H}(x)=\langle xD, \{xD\setminus M_1, \dots, xD\setminus M_k\}\rangle$. 
\end{thm}

\begin{exam}
 Consider the target $t\in A$, for Table \ref{Tab13*}. Removing $[t,1_L]$ from $L(T)$ in Figure \ref{fig:lattice6}, we get a poset with three maximal elements: $a_1$, $s$ and $b_1\vee b_2$. Seeing the same lattice as a closure lattice of operator $\phi_A$, these are closed sets $M_1=\{a_1, b_1$\}, $M_2=\{s,a_2,b_2\}$ and $M_3=\{b_1,b_2\}$. These elements can be easily found directly in the table: they correspond to $u\in U$ such that $(u,t)\in \uparrow$; in the case of Table \ref{Tab13*}: $u=1,3,4$ and $M_1=sup_U(1)$, $M_2=sup_U(3)$, $M_3=sup_U(4)$.
 Then the related hypergraph $\mathcal{H}(t)=\langle tD, \{tD\setminus M_1, \dots, tD\setminus M_3\}\rangle =\langle \{a_1,a_2,b_1,b_2\}, \{a_1b_1, a_2b_2, a_1a_2\}$.
 
 It is easy to find the minimal transversals: $a_1b_2, a_2b_1, a_1a_2$. As Theorem \ref{D&dual} claims, the left sides of both $D$-basis implications --
$a_1b_2\to t, a_2b_1\to t$ -- are present among minimal transversals.

\end{exam}

Note that Theorem \ref{D&dual} only states that every implication $Y\to x$  from the $D$-basis will be accounted for among the minimal transversals of a specific hypergraph $\mathcal{H}(x)$. On the other hand, there might be transversals that do not represent an implication from the $D$-basis, but they necessarily represent implications from a larger basis known as the \emph{canonical direct unit basis}; see \cite{Ber2010}. 

The paper \cite{Ada2025} provides a new version of the algorithm for the exact retrieval of the $D$-basis based on the dualization in distributive lattices \cite{Elb2022}. It could support a new code implementation of the $D$-basis in the future. 

We note that, in principle, the discovery of implications $Y\to x$ in table $T$ discussed in Theorem 1, can be considered as a type of ``subgroup discovery", a line of research within FCA community, see, for example, \cite{Kat18,Li16}. 

For both the \emph{Small Space}, introduced in section 1,  and the current implementation of the $D$-basis, the time complexity is defined by the time complexity of the hypergraph dualization stage. In the seminal paper \cite{FK96} the time complexity of the algorithm for the dualization of a hypergraph was shown to have a sub-exponential upper bound based on the combined size of the input and output. The question of whether there exists such an algorithm with polynomial time complexity remains open. Multiple algorithms for dualization are presented and analyzed in \cite{MU14}. 

In particular, the Reverse Search (RS)
algorithm relies on the concept of a critical hyperedge to guide the search for minimal transversals without needing to check its output against a list of transversals already found, which makes the search effective in both time and space while still finding every minimal transversal only once. 

These characteristics make the RS algorithm ideal when moving the ``update frequencies" functionality into the hypergraph dualization process, as hyperedges can be discarded rather than stored without affecting the enumeration behavior of the RS algorithm. The next section gives more details on the essence of this change and its importance for reducing memory usage in data analysis applications.

\vspace{0.3cm}

\section{Ranking attributes relevant to a target attribute}\label{A:rank}

In this section we outline the ``accounting" function that the current $D$-basis implementation performs. We note that this particular function does not need to deal with implications of the $D$-basis, and instead can be applied to any set $\mathcal{T}$ of association rules or implications.
When such a set $\mathcal{T}$ is available, one can measure the frequency of any attribute $y$ appearing  
in the antecedents of the rules $Y\to t$ in  $\mathcal{T}$, together with other attributes. 
One advantage of such an approach is that the typically high volume of rules found in the table 
provides a good representation of all attributes and allows for better comparison of attributes related to a given target.

For the purposes of the $D$-basis algorithm, $\mathcal{T}$ is one of two sets. One is $\mathcal{T}(t, \delta)$, the set of implications $Y\to t$ in the $D$-basis, with some fixed target $t \in A$ and value $\delta > 0$ so that $|\sup_A(Y\cup t)| \geq \delta$. Here $\delta$ is the \emph{minimal support}, which says that, for every implication in $\mathcal{T}(t, \delta)$, there exist at least $\delta$ rows of table $T=(U,A,R)$ that support it. 
 
The second set $\mathcal{T}(\neg t,\delta)$ is similar, except $t$ is replaced by $\neg t$, which may or may not be a member of $A$. When $\neg t$ is not in the table, one needs to modify the table by adding $\neg t$ (or replacing $t$ with $\neg t$) before running the algorithm.

Let us give a more precise definition of how the relevance of attribute $y$ with respect to target attribute $t$ is computed. 
For each attribute $y \in A\setminus{t}$ and some $\delta > 0$, the \emph{relevance} of $y$ to $t \in A$ is a ratio of two parameters of \emph{total support}. One is computed with respect to set $\mathcal{T}(t,\delta)$ of implications describing the table:

\begin{equation}
\label{eq:tsup}
\tsup_t(y)=\Sigma \{\frac{|sup_A(Y)|}{|Y|}
:  y\in Y, (Y\rightarrow t)\in \mathcal{T}(t, \delta)
\} .
\end{equation}

Thus $\tsup_t(y)$ shows the frequency of attribute $y$ appearing together with some other attributes in implications $Y\rightarrow t$ of the basis $\mathcal{T}(t,\delta)$. The contribution of each implication $Y\rightarrow y$, where $y \in Y$, to the total support computed for $y$ is higher when $\sup_A(Y)$ is higher, 
but also when the antecedent $Y$ has fewer other attributes besides $y$.\\

\begin{exam}
 The following data set has been converted to a binary form from a snippet of real data from 10 patients with two types of liver cancer. This data was tested during the initial implementation of the $D$-basis algorithm leading to publication \cite{ANO15}.

\vspace{0.3cm}
\begin{center}
1\ 0\ 0\ 0\ 0\ 1\ 0\ 0\ 0\ 1\ 1\ 0\ 1\ 1\ 0\ 1\ 1\ 1\ 0\ 1\ 1\ 1\\ 
1 1 1 0 0 1 0 0 1 1 1 0 0 0 1 1 0 1 0 1 0 1\\
1 1 1 1 0 1 1 1 1 1 1 1 1 0 0 0 0 0 0 0 0 1\\
1 1 0 1 0 1 1 1 0 0 0 0 1 1 1 0 0 0 1 1 1 1\\ 
0 0 1 1 1 1 1 1 1 1 1 1 1 1 1 1 1 1 1 0 0 0\\
0 0 0 1 1 1 1 1 0 0 1 1 0 0 1 1 0 0 0 1 1 0\\
1 1 1 1 1 1 1 1 0 1 0 1 0 1 0 1 0 1 0 1 0 1\\ 
1 1 0 1 0 1 1 1 1 1 1 0 1 1 1 1 1 1 0 1 1 1\\ 
1 1 0 0 0 1 1 1 0 0 0 0 0 0 0 1 1 1 1 0 0 0\\
0 0 0 0 1 1 1 1 0 0 0 0 1 1 1 1 0 0 0 0 1 1\\
 \end{center}
 
The set of implications $\mathcal{T}(22,3)$ was requested with the target attribute $t=22$, which is the last column in the table, and $minsup=3$.  The algorithm produced 17 implications with the consequent $t=22$ and support ranging between 3 and 6. The output includes information about reduced columns and lists all implications with their supports and the list of rows where each implication is validated:\\
\\
$6\Leftrightarrow $ \ \ \ \ \ \ \ Note: column 6 is reduced, a column with all 1s\\
$8 \Leftrightarrow7$ \ \ \ \ \  Note: column 8 is reduced, equal to column 7\\

1;\ 1 4 $\rightarrow$ 22 ; Support = 4; rows = 3, 4, 7, 8,\\

2;\ 1 10 $\rightarrow$ 22 ; Support = 5; rows = 1, 2, 3, 7, 8,\\

3;\ 1 11 $\rightarrow$ 22 ; Support = 4; rows = 1, 2, 3, 8, \\

4;\ 1 21 $\rightarrow$ 22 ; Support = 3; rows = 1, 4, 8,

...\\

17;\ 21 14 $\rightarrow$ 22 ; Support = 4; rows = 1, 4, 8, 10\\

The last part of the output provides the total supports for all attributes that participate in produced set of implications $\mathcal{T}(22,3)$, computed according to equation (\ref{eq:tsup}):
\begin{table}[H]
\begin{center} 
\begin{tabular}{|c|c|c|c|c|c|c|c|c|c|c|c|c|c|c|c|c|} 
  \hline
  Column $y$ & 1 &  ... & 4 & ... & 10 & 11 &12 & 13 & 14 & 15 & ... &  18 & 19 & 20 & 21 & 22\\
  \hline
  $tsup_{22}(y)$ & 10.67 &0  & 2&0&3& 2 & 0 & 3.67 & 4&1&0& 1.33 &0 & 6.67 & 3.67&0\\
 \hline
\end{tabular}
\label{Tab13}
\end{center}
\end{table}

For example, attribute 4 appears only in implication \#1, and its contribution to the total support is given by: $$\frac{Support(1\ 4 \rightarrow 22)}{|\{1,4\}|}=2.$$

\end{exam}

While the frequent appearance of a particular attribute $y$ in implications $Y\rightarrow t$ might indicate the relevance of $y$ to $t$, the same attribute may appear in implications $U\rightarrow \neg t$.

Let $\mathcal{T}(\neg d, \delta)$ be the basis of closure system obtained after replacing the original column of attribute $t$ by its complement column $\neg t$. Then the \emph{total support} of $\neg t$ can be computed, for each $y \in A\setminus t$, as before:
\[
\tsup_{\neg t}(y)=\Sigma \{\frac{|sup(U)|}{|U|}
%* conf(X\to \neg d)
:  y\in U, (U\rightarrow \neg t)\in \mathcal{T}(\neg t, \delta))\} .
\]

Now define the parameter of \emph{relevance} of attribute $y \in A\setminus t$ to attribute $t$, with respect to the pair of bases $\mathcal{T}(t, \delta)$ and $\mathcal{T}(\neg t, \delta)$: 
\begin{equation}
\label{eq:rel}
\rel_t(a) = \frac{\tsup_t (y)}{\tsup_{\neg t} (y) +1}.
\end{equation}

A high relevance of $y$ to $t$ is achieved by a combination of high total support of $y$ for implications $Y\rightarrow t$ and low total support for implications $U\rightarrow \neg t$.
This parameter provides a ranking of all attributes $y \in A\setminus t$.\\

Note that the ranking of all attributes in relation to a particular target attribute was introduced in 
\cite{ANO15}, together with the first application of the $D$-basis algorithm in data analysis. It was later successfully applied for analysis of survival probability in patients with stomach cancer in \cite{NCSLA21}, when the number of attributes (gene expressions) was reduced to about 500 by other methods. There was no direct analysis of the implications found, due to the sheer size of the output sets.

This aspect of implications and, more generally, association rules found in attributes of binary data is discussed in \cite{Kry2002, Bal2010}, with the goal of restricting the attention to concise subsets of rules that would also allow some logical inference.

Instead, the approach with the ranking proposed in \cite{ANO15} relies on a statistical treatment of rules, where a particular statistic called the \emph{relevance} given by equation (\ref{eq:rel}) indicates that a particular attribute may have a higher or lower probability of appearing in implications with the given target.

From the point of computational effectiveness, this approach allows the algorithm to keep less information in the output: the total support values for each attribute can be returned alone, rather than the whole set of implications.

\subsection{Small Space version of the $D$-basis algorithm}

In this section we describe the flow of the current version of the $D$-basis algorithm, then demonstrate the modifications made to this flow in the Small Space version.

The code and documentation of the $D$-basis algorithm is available through the public Git repository \url{https://gitlab.com/npar/dbasis}.

\begin{figure}[H]
    \centering
    \includegraphics[scale=0.36]{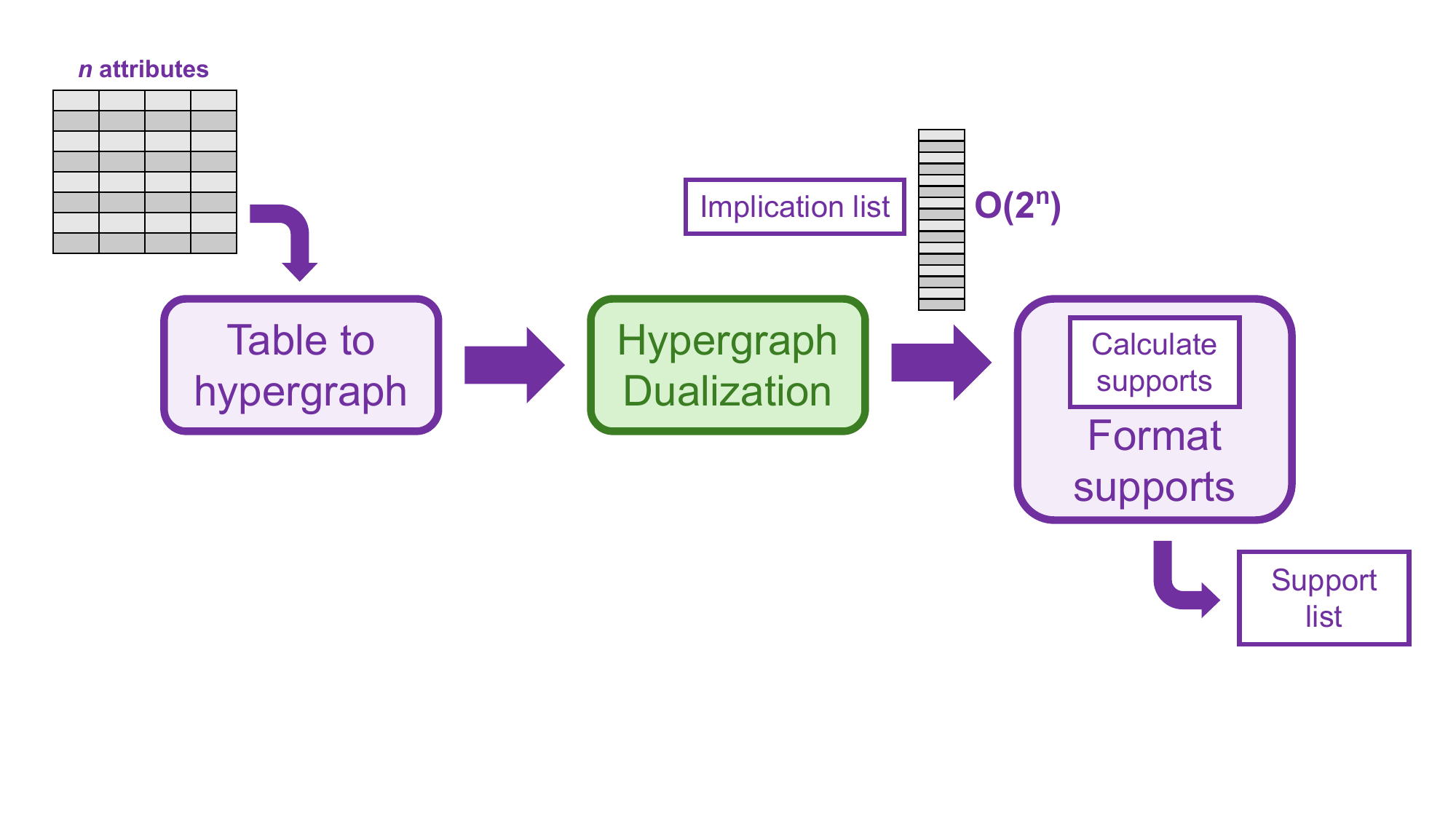}
    \caption{Flow of the $D$-basis algorithm}
    \label{fig:Dbas}
\end{figure}

Referring to Fig.\ref{fig:Dbas}, here are the main phases of the algorithm:
\begin{enumerate}
    \item The algorithm is initialized with the binary table, together with several parameters: the index of the target column $t$, the minimum support $minsup$ of implications accepted for the output, and the name of the specific implementation of hypergraph dualization (HD) to be used by the algorithm.
    \item The table is reduced, the $D$-relation for $t$ is computed, and the hypergraph $\mathcal{H}(t)$ is passed to the specified HD library.
    \item The HD library produces the dual hypergraph $\mathcal{H}^d(t)$, whose hyperedges are passed back to the main unit.
    \item Hyperedges returned by HD are used to form implications with $t$ as a consequent and checked against the original table to ensure they pass the $minsup$ threshold. Implications which do not have a support of at least $minsup$ are discarded.
    \item For every implication $Y\to t$ which meets the $minsup$ requirement, its contribution to the total support of each attribute appearing in $Y$ is calculated, and an array of total supports is updated with these contributions.
    \item The output includes the list of all implications and the total supports array.
\end{enumerate}

\begin{figure}[H]
    \centering
    \includegraphics[scale=0.36,page=2]{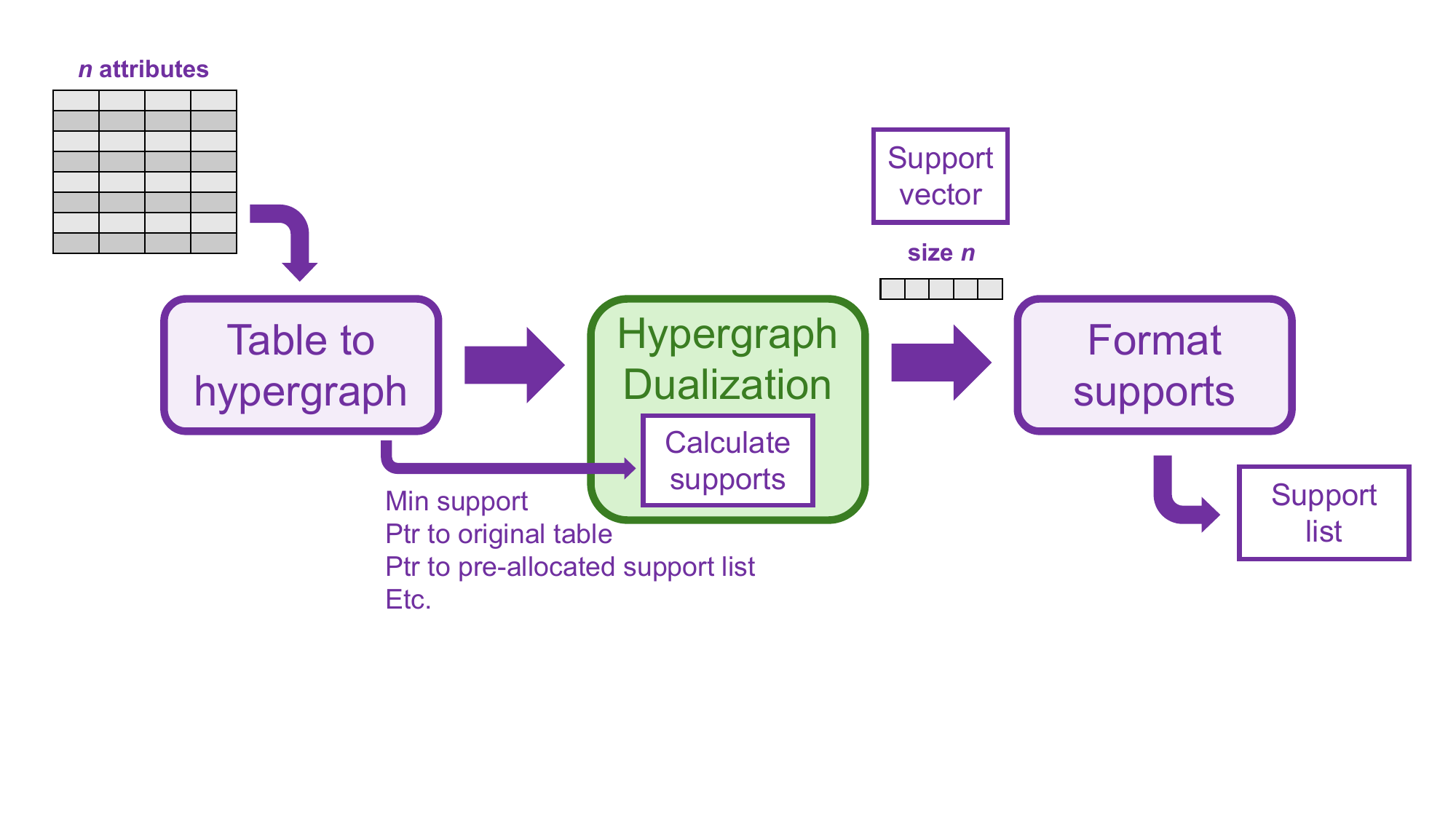}
    \caption{Small Space $D$-basis algorithm}
    \label{fig:Smsp}
\end{figure}

Fig. \ref{fig:Smsp} shows the flow of the modified version of the algorithm, called the Small Space $D$-basis.

\begin{itemize}
    \item The algorithm starts similarly, being initialized with the binary table, the index of the target column $t$, and the minimum support $minsup$ of implications for the output. The choice of HD algorithm is restricted to RS (Reverse Search).
    \item The table is reduced, and the $D$-relation for $t$ is computed. The hypergraph $\mathcal{H}(t)$ is passed to the RS algorithm in the HD library, along with pointers to original representation of the binary table and an empty array allocated for total supports.
    \item The RS algorithm generates the dual hypergraph $\mathcal{H}^d(t)$ one transversal at a time. When a minimal transversal $Y$ is found, the implication $Y\to t$ is immediately tested on the original table to ensure it passes the $minsup$ threshold. If the $minsup$ threshold is passed, its contribution to the total support of each attribute in $Y$ is calculated, and the total supports array is updated. The hyperedge $Y$ is then discarded, rather than being stored.
    \item After the RS algorithm has run, the total supports array is passed back to the main unit. As hyperedges were not stored in the previous step, no list of implications is returned by the HD library.
    \item The output includes only the total supports array.
\end{itemize}

By calculating the total support contribution of each edge as it is found during HD, the Small Space algorithm bypasses the need to store the entire list of hyperedges in $\mathcal{H}^d(t)$ at once. As this list has size complexity $O(2^n)$ relative to the number of attributes in the input table, it is easily the fastest-growing source of memory consumption in the $D$-basis algorithm. Therefore, in data analysis applications on large datasets (\cite{stem2025}, \cite{ABLS23}, \cite{NCSLA21}), where only the total supports are required to identify relevant attributes, eliminating the need to store the implication list can dramatically reduce the algorithm's maximum memory usage on larger datasets.

\section{Testing results}

Here we present the results of the Small Space testing for several data sets. The open-source program Valgrind (\url{https://valgrind.org/}) was used to track memory usage, instruction count, and program runtime over the duration of the algorithm's run. Specifically, we used the Massif tool, which is a heap profiler provided by Valgrind to track the memory use over total memory use, time, and instruction use as separate tests.

Tests described in sections \ref{sec:stem} and \ref{sec:cips} were performed in a Docker container, hosted on a laptop computer with an AMD Ryzen 7 5800H processor. For each column $t$ of the table, Valgrind was used to run the original $D$-basis algorithm with target $t$ three times, measuring the maximum memory utilization of the first run, the number of instructions executed during the second run, and the runtime (in milliseconds) of the third run. All test runs used a minimum support of 1, so vacuously true implications were excluded, but all implications supported by at least one row of the input table were permitted. This same process was then repeated using the Small Space algorithm. Results were recorded for all six runs and compared against one another to evaluate the impact of the Small Space's changes.

The tests in section \ref{sec:young} were tested on the StarHPC cluster of Hofstra University.

The data sets and instructions on how to use the $D$-basis code and read the output are given in the repository: 
\url{https://gitlab.com/npar/dbasis/-/tree/pipeline_testing}

\subsection{STEM dataset}\label{sec:stem}

The first testing dataset was taken from the data involved in publication \cite{stem2025}, where the first and third authors were involved in as co-authors. This data comprises the records of about 100 undergraduate students in academic years 2010-2015, who majored in Biology at a private institution of higher education on Long Island, NY. Each student's record had binary-type information that indicated their grades in mathematics, biology and chemistry classes in their freshman year, as well as their choice of mathematics class; their GPA in the first two semesters; whether they were retained by the school after one year; whether they switched their major; their time and place of graduation, if any, including whether they transferred before graduating; and the number of attempted and completed credits. 

The aim of the study was to determine which attributes of a student's data would be highly relevant to the likelihood of the student graduating; and among graduated students, which attributes were relevant to graduating within 4 years versus in 5 or more years, or to graduating within a {\bf STEM} ({\bf S}cience {\bf T}echnology {\bf E}ngineering and {\bf M}athematics) field versus changing major to a non-STEM field. Thus, the tests were done targeting a few 'outcome' attributes, such as degree major and time to graduation. The article \cite{stem2025} then proposes practical steps based on the findings which can be adopted by universities to potentially improve graduation rates.

The total number of binary attributes in the data is 220, thus, the binary entry table for the $D$-basis algorithm is of relatively large size $88\times 220$. Because many of the columns represent a binary encoding of attributes with large ranges of possible results, the density of 1s within the binary table is only 0.14. Tests were only run with target columns 1 through 30, but all columns were present in the input table.

The results for every target column (maximum memory usage, processor instruction count and runtime of both implementations) are given in the files \texttt{stem\_memory\_bytes.csv}, \texttt{stem\_instructions.csv} and \texttt{stem\_runtime\_ms.csv} of the directory \texttt{paper\_results} in the repository, together with a description of their columns: \url{https://gitlab.com/npar/dbasis/-/tree/pipeline_testing/paper_results}.

For 7 out of the 30 target columns, the target was found to be reducible to (an)other column(s) in the table. In both implementations of the algorithm, this halts execution before hypergraph dualization is called, resulting in nearly identical results between the two versions. On the remaining 23 target columns, however, the Small Space implementation showed significant memory savings over the original version, having an average maximum memory utilization across those 23 columns of only 441 KB compared to the original $D$-basis's average of 11.70 MB—a 93.9\% average reduction. Across those same 23 target columns, the Small Space implementation saw an increase in the number of processor instructions by an average of 27.2\% of the original $D$-basis's instruction count, while the average runtime for each column decreased by 23.2\% of the original.

\subsection{Impostor Phenomenon dataset}\label{sec:cips}

Another dataset included in testing involved the standard survey on Impostor Phenomenon (IP), collected for an ongoing analysis project by a collaborator at Queensborough Community College (QCC), Dr. R. Nelson.
The impostor phenomenon, or the persistent belief that one’s success results from luck, mistakes, or others’ misperceptions rather than true ability, has become a key concept for understanding motivation and well-being in higher education \cite{CI1978}. The 20-item Clance Impostor Phenomenon Scale (CIPS) is the most commonly used tool for measuring IP in a person, with score ranges indicating how severe a person’s impostor feelings are. Data collected includes demographic information including age, gender, race/ethnicity, office hours attendance, and tutoring center attendance, as well as a 20-question survey from Clance which yields Likert Scale responses (1-5).

The data set collected by R. Nelson included the survey answers of 84 students in Biology classes at QCC. The initial analysis of these responses are reported in \cite{NM2025}. Another project is currently underway \cite{AKN25} which involves analysis of the implications generated by the $D$-basis, in addition to the ranking of attributes with respect to particular targets in the dataset. For that project, the data was converted into a binary format, where each question in the 20-item survey was given three binary columns: one for responses at level 4-5 (indicating intense feelings of IP expressed for that question), one for responses at level 3 (medium intensity responses), and the third for responses at level 1-2 (low intensity responses). Other attributes of the participants recorded in the binary table included their gender, age and ethnicity, as well as their participation in office hours and/or tutoring.
The goal of the study was to connect these non-survey attributes with higher or lower manifestation of IP in an individuals, given that almost no IP study had ever been conducted on the population of community colleges. 

The total input table has the size $84\times 85$, making it almost square, in contrast with the data in section \ref{sec:stem}. The density of 1s within the binary table is 0.31. Tests were run with target columns 1 through 82.

The results for every target column are given in the files \texttt{cips\_memory\_bytes.csv},\\ \texttt{cips\_instructions.csv} and \texttt{cips\_runtime\_ms.csv} of the same directory \texttt{paper\_results}.
Of the 82 target columns, only one (column 70) was found to be reducible to (an)other column(s) in the table. On the remaining 81 target columns, the Small Space implementation once again showed significant memory savings, a small increase in average instruction count, and a decrease in average runtime. The average maximum memory utilization across the 81 successful target columns was 280 KB, compared to the original $D$-basis's average of 34.14 MB, for an average reduction of 98.9\%. The number of processor instructions executed by the Small Space implementation increased by an average of 9.3\% from the original, while the runtime per target column decreased by an average of 31.6\% of the original. 

Despite fewer attributes being present in this dataset than the STEM data (see \ref{sec:stem} above), the memory and runtime savings of the Small Space implementation were even more pronounced, and the instruction count increase less severe. This discrepancy is due to the average number of relevant implications produced by each dataset, which stems from the density of 1s in each dataset's binary table. With similar row counts and identical minimum support thresholds, the higher density of 1s in the Impostor Phenomenon data likely produced more implications than the STEM data, resulting in a longer processing time and a larger list of implications for the original implementation to store. 

\subsection{Young People dataset}\label{sec:young}

  \textbf{Data source.} The third dataset is the Young People Survey, published on Kaggle (\url{https://www.kaggle.com/datasets/miroslavsabo/young-people-survey/data}). According to its description, the survey was
  conducted in 2013 by the students of the Statistics class at FSEV UK, who asked their friends to take part. The questionnaire was originally in Slovak and was later translated into English; it was offered in
  electronic and in written form, and all participants were of Slovak nationality and between 15 and 30 years old. The data file \texttt{responses.csv} has 1010 rows and 150 columns (139 integer and 11 categorical) and
  contains missing values. The variables cover music preferences (19 items), movie preferences (12), hobbies and interests (32), phobias (10), health habits (3), personality traits, views on life and opinions (57),
  spending habits (7), and demographics (10).

  \textbf{Binary tables.} Most fields are Likert-scale responses (1--5). Each such field was replaced by binary columns for a low (1--2), medium (3) and high (4--5) response and for a blank (missing) response; for
  example, \texttt{Classical music} became \texttt{Classical\_music\_l}, \texttt{Classical\_music\_m}, \texttt{Classical\_music\_h} and \texttt{Classical\_music\_b}. Categorical fields were encoded in the same way,
  e.g., the answers to the question on punctuality became the columns \texttt{Punctuality\_early}, \texttt{Punctuality\_timely} and \texttt{Punctuality\_late}. Columns with too few 0s or too few 1s were then removed,
  which gave the table \texttt{filtered\_output.csv}. From this table, nine tables were generated by taking the first 50, 100 or 200 rows of \texttt{responses.csv} and 50, 100 or 200 semi-randomly selected columns;
  below, the table with $r$ rows and $c$ columns is denoted $r\times c$. The tables are nested: every table consists of the first rows and the first columns of each table that is at least as large in both dimensions.
  For example, the $50\times 100$ table extends the $50\times 50$ table by 50 new columns, and the $100\times 50$ table extends it by 50 new rows. The density of 1s in the nine tables is between 0.43 and 0.45. The
  tables (files \texttt{binary\_table\_<rows>\_<columns>.dbm}), the instructions for converting them into the input format of the $D$-basis program (\texttt{instructions.md}) and the script that runs the Valgrind tests
  (\texttt{script.sh}) are in the directory \texttt{memtest\_data}.

  \textbf{Tests on the StarHPC cluster.} The nine tables were tested on the StarHPC cluster of Hofstra University. These tests had three goals: to repeat the measurements on dedicated hardware, isolated from other
  workloads and with enough repetitions to report averages and their spread; to obtain the memory measurements for the two largest tables, which could not be completed with Valgrind before; and to examine why the Small
  Space implementation can execute more instructions than the original one and still finish sooner.

  \textbf{Setup.} The tests ran on compute nodes with two AMD EPYC 7513 32-core processors and 1 TB of RAM, under Red Hat Enterprise Linux 8.10 and the Slurm 21.08 workload manager. Both implementations were compiled
  from the public repository with the same toolchain (GCC 11.2.0 with \texttt{-O3}, Boost 1.87.0): the original $D$-basis from branch \texttt{master} (commit \texttt{45f7f1b}) and the Small Space implementation from
  branch \texttt{pipeline\_testing} (commit \texttt{2a75a35}). 
  %The only change to the source code was a missing \texttt{\#include <stdexcept>} added to the original version, so that it compiles with GCC 11. 
  Every run
  used the options \texttt{-c 2 -mins 2}, as in \texttt{script.sh}: the target is column 2, \texttt{Slow\_songs\_or\_fast\_songs\_m}, which is the same column in all nine tables, and the minimum support parameter is 2.

  \textbf{Patching Valgrind.} Valgrind executes the profiled program inside its own address-space manager. On 64-bit Linux, Valgrind 3.26.0 allows the program at most 128 GiB of address space; this limit is the
  constant \texttt{aspacem\_maxAddr} in the file \texttt{coregrind/m\_aspacemgr/aspacemgr-linux.c}. The original implementation exceeds this limit on the $200\times 200$ table, and under unmodified Valgrind its
  profiling run terminated when the virtual memory of the job reached 127.9 GiB. We changed this constant from \texttt{0x2000000000 - 1} (128 GiB) to \texttt{0x40000000000 - 1} (4 TiB). The Valgrind build we used also
  had the maximum number of address-space segments (\texttt{VG\_N\_SEGMENTS} in the same file) raised from 30,000 to 500,000. Valgrind was rebuilt from source with these changes. They only raise the amount of memory
  the profiled program may use and do not change how Massif measures: for the eight smaller tables, the peaks measured by the original implementation are identical to those obtained earlier with the unmodified
  Valgrind. With the patched Valgrind, the profiling of the original implementation on the $200\times 200$ table completed successfully after 2 days and 2 hours, on a node with 2 TB of RAM; Valgrind itself occupied
  about 1 TB (1009 GiB) of resident memory during this run.

  \textbf{Memory measurement.} Peak heap usage was measured with the Massif tool of the patched Valgrind (option \texttt{--time-unit=B}, as in \texttt{script.sh}). Each version was profiled 10 times per table, except
  for the $200\times 200$ table, which was profiled once per version. The peak of the original implementation was the same in all 10 runs; for the Small Space implementation, the standard deviation was at most $5.9\%$
  of the mean.

\textbf{Runtime and instruction measurement.} The runtime was measured natively, without Valgrind. For each run, a whole node was reserved, so no other jobs shared it. Each table was run 10 times per version, on a
  single thread pinned to one physical core. The order of the two versions alternated between repetitions. Wall-clock, user and system time, maximum resident memory and page faults were recorded with
  \texttt{getrusage}, and user-space hardware counters (instructions, cycles, last-level cache misses) with the Linux \texttt{perf\_event\_open} interface. The standard deviation of the wall-clock time was at most
  $2.4\%$ of the mean for every table and version. The two smallest tables ($50\times 50$ and $100\times 50$) run in milliseconds, and their timings were taken from a separate series that began with a warm-up run of
  each version: in the main series, the first runs after the original $200\times 200$ run were delayed by reloading the program files from the network file system.

  \begin{table}[H]
      \centering
      \begin{tabular}{|l|r|r|r|r|r|}
      \hline
       & & \multicolumn{2}{c|}{\textbf{Peak heap (Massif)}} & \multicolumn{2}{c|}{\textbf{Wall time (s)}} \\ \hline
          \textbf{Table} & \textbf{Min. transversals} & \textbf{Original} & \textbf{Small Space} & \textbf{Original} & \textbf{Small Space} \\ \hline
          50 x 50   & 7,853         & 1.9 MiB  & 0.65 MiB & 0.0291 & 0.0203 \\ \hline
          50 x 100  & 239,389       & 53.7 MiB & 1.27 MiB & 0.631  & 0.412  \\ \hline
          50 x 200  & 9,658,841     & 2.2 GiB  & 2.40 MiB & 25.4   & 18.5   \\ \hline
          100 x 50  & 22,152        & 5.2 MiB  & 0.83 MiB & 0.0868 & 0.0617 \\ \hline
          100 x 100 & 1,583,314     & 370 MiB  & 1.47 MiB & 5.69   & 4.11   \\ \hline
          100 x 200 & 94,737,343    & 22.3 GiB & 2.71 MiB & 345.4  & 289.9  \\ \hline
          200 x 50  & 74,083        & 17.6 MiB & 1.17 MiB & 0.370  & 0.305  \\ \hline
          200 x 100 & 12,457,516    & 3.0 GiB  & 2.04 MiB & 66.4   & 55.3   \\ \hline
          200 x 200 & 1,438,284,502 & 357 GiB  & 3.42 MiB & 8,689  & 7,879  \\ \hline
      \end{tabular}
      \caption{Young People tables on the StarHPC cluster: number of minimal transversals found by the dualization, peak heap usage measured by Massif (mean of 10 runs; $200\times 200$: one run), and native
  single-thread wall-clock time (mean of 10 runs).}
      \label{HpcResults}
  \end{table}

\vspace{0.5cm}

\begin{table}[H]
      \centering
      \begin{tabular}{|l|r|r|r|}
      \hline
          \textbf{Table} & \textbf{Original} & \textbf{Small Space} & \textbf{Change} \\ \hline
          50 x 50   & 0.227  & 0.149  & $-34.5\%$ \\ \hline
          50 x 100  & 6.54   & 4.72   & $-27.8\%$ \\ \hline
          50 x 200  & 260.5  & 220.7  & $-15.2\%$ \\ \hline
          100 x 50  & 0.786  & 0.582  & $-26.0\%$ \\ \hline
          100 x 100 & 54.6   & 46.2   & $-15.3\%$ \\ \hline
          100 x 200 & 3,311  & 3,413  & $+3.1\%$  \\ \hline
          200 x 50  & 3.32   & 2.91   & $-12.3\%$ \\ \hline
          200 x 100 & 582.9  & 606.5  & $+4.0\%$  \\ \hline
          200 x 200 & 68,367 & 88,172 & $+29.0\%$ \\ \hline
      \end{tabular}
      \caption{Young People tables on the StarHPC cluster: user-space instructions executed (in billions, $10^9$), measured with hardware performance counters in native single-thread runs (mean of 10 runs). Change $=$
  (Small Space $-$ Original)/Original.}
      \label{HpcInstructions}
  \end{table}

  \textbf{Memory.} The results are shown in Table \ref{HpcResults}. The peak heap usage of the original implementation is proportional to the size of its output: it uses between 235 and 266 bytes per minimal
  transversal on all nine tables, reaching 357 GiB (490 GiB including the overhead of the memory allocator) for the 1.44 billion minimal transversals of the $200\times 200$ table. The peak of the Small Space
  implementation does not depend on the number of transversals and stays between 0.65 MiB and 3.42 MiB. The memory gain therefore grows with the output: from a factor of 3 for the $50\times 50$ table to a factor of
  about $107{,}000$ for the $200\times 200$ table.

  \textbf{Runtime.} The Small Space implementation was faster on every table, by a factor of 1.10 to 1.53. For both implementations, the runtime grows with the number of minimal transversals: among tables with the same
  number of rows, the time per minimal transversal changes by at most a factor of 1.5 for either version, while the number of transversals in Table \ref{HpcResults} ranges from 7,853 to 1.44 billion. This agrees with
  both implementations having the same asymptotic complexity; the Small Space implementation reduces the time per transversal.

  \textbf{Instructions.} The instruction counts are shown in Table \ref{HpcInstructions}. For every table and version, the count was the same in all 10 runs. The counts do not depend on the measuring tool: for the
  $100\times 100$ and $200\times 100$ tables, the hardware counters agree within $1.2\%$ with the instruction counts obtained under the Callgrind tool of Valgrind. The Small Space implementation executed fewer
  instructions than the original on six of the nine tables, and more on the three tables with the largest outputs ($3.1\%$, $4.0\%$ and $29.0\%$ more for the $100\times 200$, $200\times 100$ and $200\times 200$
  tables).

  \textbf{More instructions, less time.} The effect observed in the earlier tests, more instructions but a shorter runtime, therefore also appears on the dedicated nodes, so it is not caused by the test environment.
  Profiling the $200\times 100$ table with Callgrind shows where the instructions go. The core of the Reverse Search dualization executes exactly the same work in both versions (e.g., $32.4\cdot 10^{9}$ instructions in
  the function \texttt{vertex\_would\_violate} in each). The Small Space version spends additional instructions computing the support of every transversal on the table rows as soon as the transversal is found
  ($221.9\cdot 10^{9}$ instructions in the Reverse Search routine that now performs this check, and $105.5\cdot 10^{9}$ in counting bits). The original version instead spends $133\cdot 10^{9}$ instructions on memory
  allocation and deallocation, $65\cdot 10^{9}$ on converting the stored transversals into implications, and further instructions on printing them. These instructions, however, do not account for the whole runtime. For
  the $200\times 200$ table, the original implementation spent 960 s in the operating system kernel and caused 717 million page faults while filling its memory, compared with 0.2 s and about one thousand page faults
  for the Small Space implementation. The original also had 6 times more last-level cache misses ($66.7\cdot 10^{9}$ vs.\ $11.0\cdot 10^{9}$) and executed fewer instructions per cycle (2.45 vs.\ 3.10). User-space
  instruction counts include neither the kernel time nor the processor cycles spent waiting for memory, so a higher instruction count does not imply a longer runtime.

  \textbf{Correctness.} Both implementations produced identical total supports for every table, in all native runs and in all Valgrind runs.
\section{Discussion}\label{sec:discussion}

In this section we summarize the results of Section 4 on the three datasets and explain what they mean. Table \ref{SummaryResults} collects the main numbers.

\begin{table}[H]
      \centering
      \small
      \setlength{\tabcolsep}{4pt}
      \begin{tabular}{|l|r|r|r|r|r|}
      \hline
       & & \multicolumn{2}{c|}{\textbf{Peak memory}} & \multicolumn{2}{c|}{\textbf{Change (\%)}} \\ \hline
          \textbf{Dataset} & \textbf{Tests} & \textbf{Original} & \textbf{Small Space} & \textbf{Instructions} & \textbf{Runtime} \\ \hline
          STEM & 23 & 11.2 MiB & 0.42 MiB & $+27.2$ & $-23.2$ \\
          ($88\times 220$) & & (1.9--23.9) & (0.40--0.45) & ($+4.9$ to $+40.7$) & ($-27.6$ to $-18.5$) \\ \hline
          Impostor Phenomenon & 81 & 32.6 MiB & 0.27 MiB & $+9.3$ & $-31.6$ \\
          ($84\times 85$) & & (4.4--56.5) & (0.24--0.28) & ($-23.7$ to $+23.6$) & ($-45.9$ to $-21.9$) \\ \hline
          Young People & 9 & 1.9 MiB to & 0.65 MiB to & $-34.5$ to & $-34.7$ to \\
          (9 tables) & & 357 GiB & 3.42 MiB & $+29.0$ & $-9.3$ \\ \hline
      \end{tabular}
      \caption{Summary of the tests. \emph{Tests}: the number of runs of the algorithm with a target that is not reducible (STEM and Impostor Phenomenon: target columns; Young People: tables $50\times 50$ to
  $200\times 200$, target column 2). For STEM and Impostor Phenomenon, the first line gives the mean over the tests and the second line, in parentheses, the range in the same units; for Young People, the range over
  the nine tables. \emph{Change} $=$ (Small Space $-$ Original)/Original, in percent: a negative change means that the Small Space implementation needed less. STEM and Impostor Phenomenon were measured under
  Valgrind on a laptop, one run per measurement; Young People natively on the StarHPC cluster, mean of 10 runs. 1 MiB $=2^{20}$ bytes.}
      \label{SummaryResults}
  \end{table}

\textbf{What is compared.} Both implementations run the same Reverse Search dualization and enumerate the same minimal transversals. They differ only in what they do with each transversal: the original implementation stores it as an implication and computes the total supports of the attributes at the end, while the Small Space implementation adds the contribution of the transversal to the total supports at once and then discards it. For the Young People tables we verified that both implementations find the same number of minimal transversals and produce identical total supports. When the target column is reducible to other columns, the algorithm stops before the dualization, and both implementations used exactly the same memory (7 target columns of the STEM dataset and 1 of the Impostor Phenomenon dataset); these targets are not included in Table \ref{SummaryResults}.

\textbf{Memory: the main gain.} The Small Space implementation needed less memory in every test of all three datasets. The memory of the original implementation grows with the size of its output, because it keeps all implications: on the Young People tables it used between 235 and 266 bytes per minimal transversal. The memory of the Small Space implementation does not depend on the output and changes little within a dataset: from 0.40 to 0.45 MiB for the STEM targets, whose original memory ranged from 1.9 to 23.9 MiB, and from 0.65 to 3.42 MiB for the Young People tables, whose number of minimal transversals ranged from 7,853 to 1.44 billion. The gain therefore grows with the output: the original implementation needed between 4.6 and 54 times more memory on the STEM targets, between 18 and 208 times more on the Impostor Phenomenon targets, and between 3 and about 107,000 times more on the Young People tables. For the largest table, $200\times 200$, the original implementation needed 357 GiB, so it could be run only on a computer with several hundred gigabytes of memory, while the Small Space implementation needed 3.42 MiB.

\textbf{Runtime: also a gain.} The Small Space implementation was also faster in every test: on average by 23.2\% on the STEM targets and by 31.6\% on the Impostor Phenomenon targets, and by 9.3\% to 34.7\% on the Young People tables. The asymptotic complexity is the same for both implementations, since both enumerate the same minimal transversals with the same algorithm; the Small Space implementation reduces the time spent per transversal. On the Young People tables, this gain was smaller on the tables with more rows (27.2--34.7\% for 50 rows and 9.3--17.6\% for 200 rows), which agrees with the fact that the Small Space implementation evaluates every minimal transversal on all rows of the table.

\textbf{More instructions, less time.} The number of executed processor instructions does not follow the runtime. The Small Space implementation executed more instructions on all STEM targets, on 69 of the 81 Impostor Phenomenon targets, and on the three Young People tables with the largest outputs, and was nevertheless faster in all of them. The extra instructions come from computing the support of each minimal transversal as soon as it is found. The original implementation, on the other hand, spends time on handling its large memory that is not counted as instructions of the program: for the $200\times 200$ table, it spent 960 s in the operating system kernel, caused 717 million page faults and had 6 times more cache misses than the Small Space implementation. The repeated tests on the dedicated nodes of the cluster show the same effect as the earlier tests on the laptop, so it is not caused by the test environment.

\textbf{Lessons learned.} First, when the goal is the ranking of attributes by their total supports, the implications do not need to be stored: computing the total supports during the Reverse Search dualization makes the memory independent of the number of implications, which grows very fast with the size of the table. Second, this saving does not cost time; the Small Space implementation was faster in all our tests. Third, the number of executed instructions is not a reliable measure of the running time of a program that uses a lot of memory; both should be measured, preferably in isolated and repeated runs. Finally, the Small Space implementation removes the memory limit but not the cost of the dualization itself: for the $200\times 200$ table, it still ran for 7,879 seconds to enumerate 1.44 billion minimal transversals, so for large tables the running time, rather than the memory, becomes the limiting factor.

\vspace{0.5cm}

{\bf \large Acknowledgements.} We are grateful to Oren Segal for his guidance in the testing phase of this project. Hofstra University provided the financial support for the authors' attendance of the AMS 2025 Fall Western Sectional meeting in Denver, Colorado, where the results of this paper were reported. 

This research was made possible in part through the use of the Hofstra Star HPC cluster, which was funded by the National Science Foundation through the Major Research Instrumentation Program (Grant No. 2320735).

\printbibliography
%{SmallSpD.bib}

\end{document}